\documentclass[reprint, showpacs, amsmath, amssymb, aps, prb, superscriptaddress]{revtex4-1}
\pdfoutput=1

\usepackage{graphicx}
\usepackage{dcolumn}
\usepackage{bm}
\usepackage{hyperref}
\usepackage{siunitx}
\usepackage{todonotes}
\usepackage{cleveref}
\graphicspath{{figures/}}

\begin{document}
	
	\preprint{APS/123-QED}
	
	\title{Suppression of charge density wave order by disorder in Pd-intercalated \texorpdfstring{ErTe$_3$}{ErTe3}}
	
	\author{J. A. W. Straquadine}
	\email{jstraq@stanford.edu}
	\affiliation{Geballe Laboratory for Advanced Materials and Department of Applied Physics, Stanford University, California 94305, USA}
	
	\author{F. Weber}
	\affiliation{Karlsruhe Institute of Technology, Institute for Solid State Physics, 76021 Karlsruhe, Germany}
	
	\author{S. Rosenkranz}
	\affiliation{Materials Science Division, Argonne National Laboratory, Lemont, Illinois 60439, USA}
	
	\author{A. H. Said}
	\affiliation{Advanced Photon Source, Argonne National Laboratory, Lemont, Illinois 60439, USA}
	
	\author{I. R. Fisher}
	\affiliation{Geballe Laboratory for Advanced Materials and Department of Applied Physics, Stanford University, California 94305, USA}
	
	\date{\today}

	\begin{abstract}
		Disorder is generically anticipated to suppress long range charge density wave (CDW) order.
		We report transport, thermodynamic, and scattering experiments on Pd$_x$ErTe$_3$, a model CDW system with disorder induced by intercalation.
		The pristine parent compound ($x=0$) shows two separate, mutually perpendicular, incommensurate unidirectional CDW phases setting in at 270~K and 165~K.
		Here we track the suppression of signatures corresponding to these two parent transitions as the Pd concentration increases.
		At the largest values of $x$, we observe complete suppression of long range CDW order in favor of superconductivity.
		We also report evidence from electron and x-ray diffraction which suggests a tendency toward short-range ordering along both wavevectors which persists even well above the crossover temperature.
		Pd$_x$ErTe$_3$ provides a promising model system for the study of the interrelation of charge order and superconductivity in the presence of quenched disorder.
	\end{abstract}
	
	\maketitle
	
	\section{Introduction}
	
	Charge order is a ubiquitous feature in the phase diagrams of many strongly correlated materials.
	In the cuprate family of high-$T_c$ superconductors, for instance, evidence of unidirectional charge-ordered states has been observed both in scattering\cite{Wakimoto2003, Hucker2011, Chang2012, Ghiringhelli2012, Comin2014, Comin2015, Jang2016, Tabis2017} and via local probes \cite{Hoffman2002, Howald2003, Wu2015}.
	In these materials, however, charge order always forms in the presence of significant disorder due to the large concentration of dopant ions or intentionally induced oxygen non-stoichiometry that are required to suppress the parent Mott-insulating state.
	Such disorder tends to frustrate the charge-ordered state, resulting in the formation of short-range-correlated domains.
	In order to fully understand what role(s) charge order and/or its fluctuations play in enhancing or suppressing superconductivity, the effects of disorder must be taken into account.
	This motivates the study of model systems which capture the essential physics of charge order without the complications of strong magnetic interactions.
	In particular, a model system which mimics the square plane symmetry of all known high-$T_c$ superconductors (cuprates and Fe-pnictides), and which hosts a unidirectional CDW state which can be affected by disorder, would be the most relevant point of comparison.
	Here, we examine the phase diagram of Pd-intercalated ErTe$_3$, a candidate system which, as we will show, fits this description.
	
	CDWs in the presence of disorder have been studied in many systems, perhaps most intensely in the transition metal dichalcogenides (TMDCs).\cite{Arguello2014, Li2017, Cho2018, Chatterjee2015, Morosan2006}
	While the physical manifestation of disorder and its effects on collective behavior all depend on microscopic details, several general statements can be drawn from this literature.
	First, disorder acts to suppress and smear CDW transitions.
	As a point of principle, even arbitrarily weak disorder prevents a long-range-ordered incommensurate CDW state \cite{Nie2014}.
	The most important consequence of this statement is that while signatures resembling the CDW phase transitions of the pristine material may still appear in the presence of disorder (likely broadened in temperature), they are, strictly speaking, only crossovers.
	Furthermore, in a system with 4-fold symmetry, a unidirectional density wave must also break a rotational symmetry by ordering along only one in-plane direction.
	Even if long-range phase coherence is precluded by disorder, a sharp phase transition which breaks this discrete symmetry ($C_4\rightarrow C_2$, Ising-like) is still allowed to occur in the presence of weak disorder.\cite{Nie2014}
	That is to say that the orientational order of the CDW may be preserved even without long-range phase coherence.
	
	Second, localized defects tend to induce Friedel oscillations in the electron density, even at temperatures far above the crossover temperature.\cite{Arguello2014}
	This quenched disorder strongly affects the growth of phase correlations of the CDW\cite{Chatterjee2015}, and thereby can alter the macroscopic properties of the material.
	
	Finally, disorder can tune the interplay between CDWs and superconductivity in subtle ways.
	For example, in the case of isoelectronic substitution in 2H-Ta(Se,S)$_2$, the CDW state is sensitive to disorder, while superconductivity is enhanced due to the increased density of available states --- in effect, superconductivity is ``protected'' from disorder according to Anderson's theorem, whereas the competing CDW state is not.\cite{Li2017}
	For electron-irradiated 2H-NbSe$_2$\cite{Cho2018} a similar trend of phase competition is seen until the CDW is suppressed completely.
	
	\begin{figure}
		\includegraphics[width=\columnwidth]{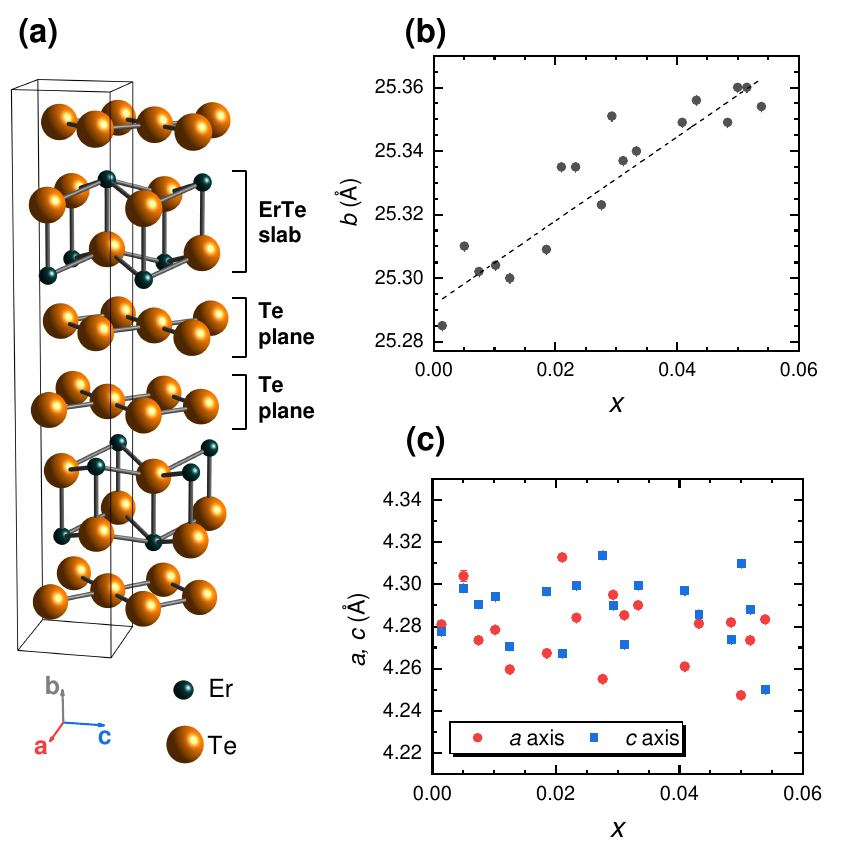}
		\caption{
			(a) Crystal structure of the $x=0$ parent compound ErTe$_3$, which consists of alternating ErTe slabs with bilayers of square Te nets.
			Unit cell is shown by black lines.
			(b) The out-of-plane lattice parameter $b$ determined by single-crystal x-ray measurements, as a function of Pd content $x$ (determined by microprobe analysis).
			Error bars are calculated as the variance in a simultaneous fit to the Cu $K\alpha_1$ and $K\alpha_2$ peaks at the (0,8,0) position.
			The lattice expands perpendicular to the planes as $x$ increases, which is consistent with incorporation of Pd atoms between the planes.
			(c) The two in-plane lattice parameters $a$ and $c$ on the same set of crystals, extracted from the (1,9,0) and (1,7,1) peaks.
			There is no significant trend in $a$ or $c$ with increasing $x$.
		}
		\label{fig:structure}
	\end{figure}

	The rare-earth tritellurides RTe$_3$ (R=Y, La-Nd, Sm, Gd-Tm) form a family of quasi-2D metals which exhibit unidirectional incommensurate CDW states. \cite{DiMasi1995, Ru2006, Ru2008, Ru2008c}
	The crystal structure is formed of alternating puckered RTe slabs with bilayers of approximately square nets of Te atoms, illustrated in \cref{fig:structure}(a).
	The presence of a glide plane in the stacking of these layers creates a 0.05\%\cite{Ru2008b} difference between the in-plane $a$- and $c$-axis lattice parameters at room temperature, and biases the primary CDW transition to order along the $c$-axis.
	Strictly speaking, the space group of RTe$_3$ is orthorhombic\cite{Norling1966} (\emph{Cmcm}, with the $b$-axis out of plane), but the electronic properties above the CDW phase transition, as witnessed for example via density functional theory calculations \cite{Laverock2005}, ARPES \cite{Moore2010} and transport measurements\cite{Ru2008b, Walmsley2016}, are effectively 4-fold symmetric.
	The Fermi surface of these materials is well described by a tight-binding model using only the Te in-plane $p_x$ and $p_z$ orbitals.
	For heavier $R$ elements (R=Tb, Dy, Ho, Er, and Tm), the material also exhibits a separate, lower-temperature phase transition\cite{Ru2008b, Hu2011a, Banerjee2013, Hu2014} characterized by a second unidirectional CDW perpendicular to the first.
	These two CDW ordering vectors appear to compete for density of states at the Fermi surface, such that suppressing one CDW by chemical\cite{Banerjee2013} or hydrostatic pressure\cite{Sacchetti2007, Lavagnini2009, Hamlin2009} enhances the other.
	High-resolution x-ray diffraction measurements on tritellurides provide a lower bound on the CDW correlation length of several microns, indicating minimal disorder effects for pristine crystals. \cite{Ru2008b}
	
	At temperatures above the CDW transition, calculations of the electronic susceptibility $\chi_q$  exhibit peaks very close to both of the orthogonal CDW wavevectors\cite{Johannes2008}.
	Similarly, measurements of the phonon dispersion\cite{Maschek2018} also shows a Kohn anomaly where the phonons at both wavevectors soften considerably when approaching the transition with decreasing temperature.
	However, the slight anisotropy induced by the glide plane causes the phonons to soften completely along the $c$-axis before the $a$-axis, instigating the CDW transition.
	Despite the proximity of $q_{CDW}$ to the peak in the calculated $\chi_q$, inelastic x-ray scattering measurements indicate that the CDW wavevector is determined by a strongly $q$-dependent electron-phonon coupling.
	
	This well-understood material family of incommensurate charge density waves may also provide a model system for exploring the evolution of unidirectional CDW correlations in the presence of disorder, which is intentionally introduced here by Pd intercalation.
	Pd atoms assume a stable, filled shell $4d^{10}$ configuration, and therefore should not affect the carrier concentration as the Pd content is increased.
	Consequently, it can be anticipated that the dominant effect of Pd intercalation is the introduction of disorder to the adjacent Te planes via local lattice strains and/or variation in the local potential.
	As we will show below, the phase diagram and physical properties of Pd$_x$ErTe$_3$ is consistent with this intuition.
	
	While many of the disorder-induced features seen in the TMDCs may be expected in the case of Pd$_x$ErTe$_3$, a systematic study is required to determine the magnitude of the effect each will have on the collective behavior of the system.
	The near-tetragonal structure and unidirectional charge ordering in RTe$_3$ set it apart from the TMDCs, which exhibit three- or sixfold crystal symmetry and often show a CDW state with three simultaneous ordering vectors oriented 120$^\circ$ from each other.
	Also, it is worth noting that all measurements point to a truly incommensurate CDW in all members of the RTe$_3$ family, with no observation of a lock-in transition to a commensurate wavevector at low temperatures. \cite{Malliakas2006, Fang2007}
	
	Polycrystalline samples of Pd$_x$RTe$_3$ (R=Y, La, Pr, Sm, Gd, Tb, Dy, Ho, Er, Tm) have been reported previously \cite{He2016}, and it was established that for high enough $x$, the material superconducts with $T_c$ $\approx$2-3~K regardless of the chosen rare earth.
	Single crystal measurements have only previously been reported in Pd$_x$HoTe$_3$ \cite{Lou2016}, and only the remnant of the first CDW transition was observed.
	From this it was assumed that the two CDWs merge together and appear simultaneously at a single crossover.
	In this work, we show transport, magnetometry, x-ray, and electron diffraction studies on samples of Pd$_x$ErT$_3$, and observe that the vestigial signatures of both of the phase transitions are suppressed independently, and that significant CDW correlations appear in both directions, even once the resistivity signature of the second CDW has been suppressed to $0~K$.
	We also demonstrate the appearance of a short-range-ordered structural distortion, evidenced by diffuse streaks in electron diffraction patterns.
	The short range correlations persist up to temperatures far above the ordering crossover, suggesting that the Pd intercalants nucleate small regions of static CDW order along both in-plane directions.
	This work establishes Pd$_x$ErTe$_3$ as a model system for studying the interplay of CDW formation and disorder close to the onset of superconductivity.
	Recent theoretical work has pointed out the potential complexities of this interplay with the expected appearance of ``fragile'' states stemming from topological defects in the CDW and vortices in the superconductor\cite{Yu2018}.
	This work sets the stage for further studies of this interplay.

	\section{Experimental Methods}
	Samples were grown using a Te self-flux as described elsewhere for pure RTe$_3$ compounds \cite{Ru2006}, with the addition of small amounts of Pd to the melt.
	Crystals of Pd$_x$ErTe$_3$ with ${0 \le x \le 0.055}$ were produced with this method, though for $x>0.03$ crystals of PdTe$_2$ also formed.
	Increasing the Er concentration in the melt and raising the temperature at which the melt is decanted minimized the formation of PdTe$_2$ in favor of Pd$_x$ErTe$_3$.
	Microprobe analysis showed that $\approx12\%$ of the Pd present in the melt incorporated into the crystals during growth, and that crystal composition is uniform to within experimental error at different spots on a crystal surface and between crystals grown in the same batch.
	Plate-shaped crystals ($b$-axis normal to the plane) 1-3~mm across were routinely produced.
	Crystal plate area remained fairly constant for all but the largest $x$ values, but the resulting thicknesses decreased as Pd concentration increased, from several hundred microns for $x=0$ to approximately 50 microns for $x=0.05$.
	This offers indirect evidence that Pd atoms act as intercalants between the Te planes, in that their presence tends to disrupt and slow the rate of growth in this direction.
	The resulting crystals are orange in color, shiny, soft, and micaceous metals.
	
	Chemical analysis was performed in a JEOL JXA-8230 “SuperProbe” electron microprobe system, calibrated to ErTe$_3$ and PdTe$_2$ secondary standards.
	
	Measurements of resistance versus temperature were performed in a Janis Supertran-VP continuous flow cryostat.
	The resistivity in the $ac$-plane was measured on thin rectangular crystals which had been cut with a scalpel and cleaved to expose a clean surface immediately before contacting.
	Crystals were cut such that current flows along the [101] axis, and contacts were attached to the surface in the transverse geometry described elsewhere. \cite{Walmsley2016}
	In this geometry, the sum of the resistivity components along the crystal axes $\rho_a+\rho_c$ and the in-plane resistivity anisotropy $\rho_a-\rho_c$ are measured simultaneously within the same crystal.
	
	Selected area electron diffraction patterns (SADPs) were taken using an FEI Tecnai G2 F20 X-TWIN transmission electron microscope.
	Samples were mounted using a liquid N$_2$ cryogenic holder with a base temperature of 95~K.
	Samples were prepared by exfoliating thin crystals, which produced large (hundreds of square microns) regions of electron transparent material free of cracks, dislocations, and significant curvature.
	While ErTe$_3$ crystals have been observed to grow with small angle twin domains as well as 90$^\circ$ stacking faults, defects such as grain boundaries and dislocations can be observed directly in TEM imaging, and are therefore avoided.
	SADPs were collected with the beam along the [010] zone axis, normal to the plane.
	
	Lattice parameter measurements were made from single-crystal x-ray diffraction in a PANalytical X'Pert PRO diffractometer using Cu $K_\alpha$ radiation.
	Comparison of superlattice peak intensity along the orthogonal directions was done on the HERIX beamline 30-ID at the Advanced Photon Source.\cite{Said2011}
	Samples were also screened for twinning and stacking faults in all x-ray measurements by comparing the intensity of the allowed (0,6,1) peak to the forbidden (1,6,0) peak prior to collecting data.
	
	Magnetic moment and susceptibility measurements were performed using a Magnetic Property Measurement System 3 from Quantum Design.
	Samples were cut into rectangular prisms, and measurements were made with field both parallel to and perpendicular to the Te planes.

	\section{Results and Discussion}
	\Cref{fig:structure}(b) and (c) show the extracted values of the lattice parameter from single-crystal x-ray diffraction measurements.
	The out-of-plane $b$-axis is observed to expand linearly with $x$, which is consistent with the Pd atoms intercalating between the square nets of the Te bilayer.
	We propose this as the most likely position for the Pd atoms, as the bilayers are only bonded by van der Waals forces.
	This observation is also consistent with the results of DFT calculations\cite{He2016} for Pd$_{0.25}$YTe$_3$.
	The in-plane lattice parameters are unchanged within experimental error.
	This argues against chemical substitution of Pd on the Er site, as the smaller radius of Pd ions would be expected to compress the lattice in-plane.
	
\begin{figure}
		\includegraphics[width=\columnwidth]{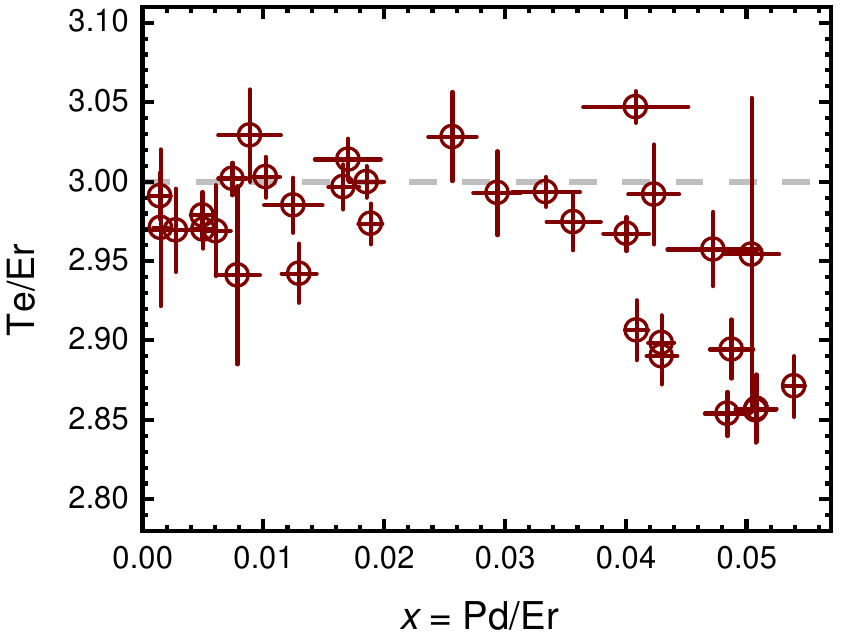}
		\caption{
			Stoichiometric ratios determined from electron microprobe measurements.
			The $x$-axis value is used throughout this work to quantify the concentration of Pd atoms per formula unit.
			We found no crystals with Pd concentrations above $x=0.055$ with this growth method.
			For growths of the highest Pd concentrations, the formation of Pd$_x$ErTe$_3$ competes with the formation of PdTe$_2$.
			For $x>0.035$, a significant deviation from the ideal ratio of 3 Te per Er is observed.
			This decrease is believed to be due to an increased population of Te vacancies within the bilayers.
		}
		\label{fig:muP}
	\end{figure}

	Results from microprobe analysis shown in \cref{fig:muP} show that Pd atoms are indeed incorporated into the material, at a rate of roughly 12\% of the initial concentration in the melt.
	This data also shows that the Te:Er ratio falls for $x>0.035$, suggesting that for large concentrations the Pd atoms displace Te atoms from the planes and generate vacancies.
	This would also be consistent with intergrowth of the RTe$_3$ structure with the related polytype R$_2$Te$_5$, in which Te bilayers alternate with single Te layers.
	However, as no extra peaks with the proper periodicity for R$_2$Te$_5$ were observed in the x-ray diffraction measurements in \cref{fig:structure}, this effect is negligible if it is present at all.
	
\begin{figure}
		\includegraphics[width=\columnwidth]{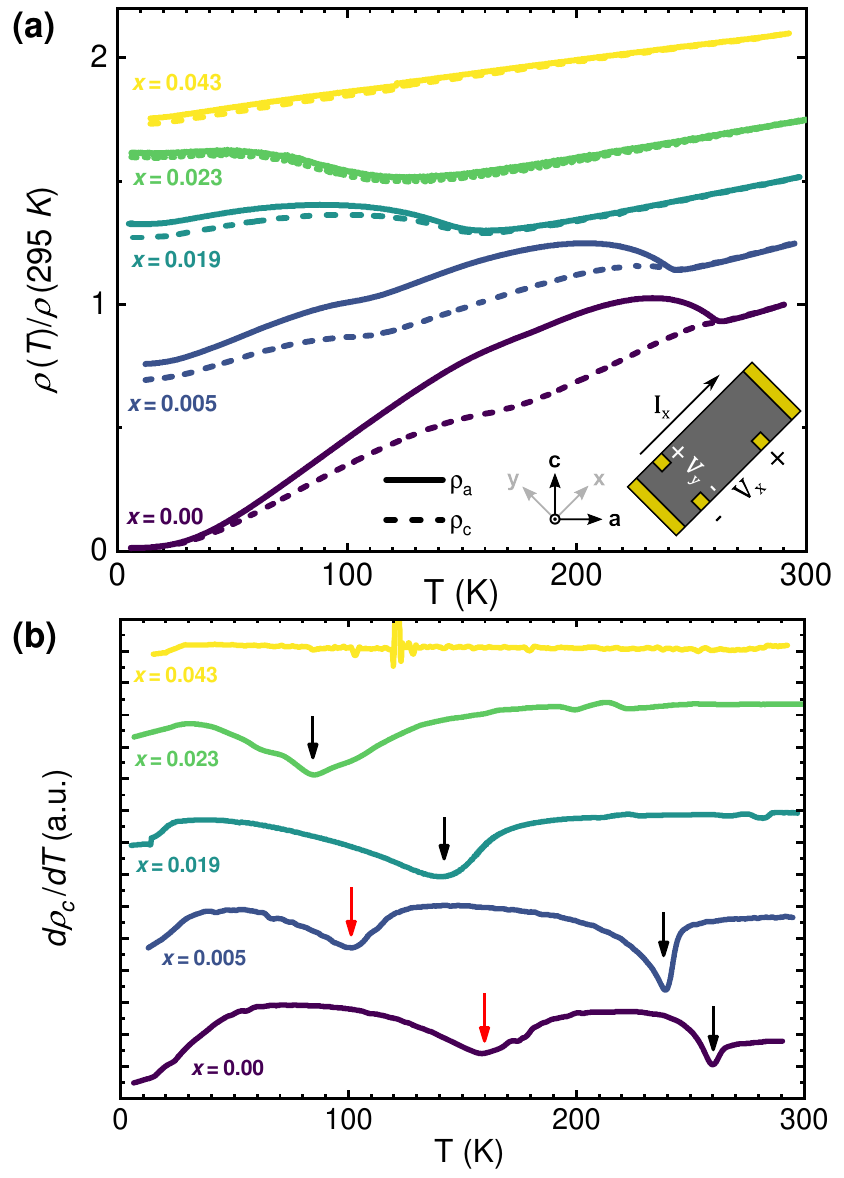}
		\caption{
			(a) Temperature-dependence of the in-plane resistivity between 4~K and 300~K, resolved into the two in-plane directions, normalized to the value at room temperature, and offset for clarity.
			Data is shown for representative compositions only.
			Solid and dashed lines correspond to the resistivity along the $a$ and $c$ axes, respectively.
			The inset describes the measurement geometry, in which a rectangular bar is cut with a scalpel at a 45$^\circ$ angle to the crystal axes.
			In a coordinate frame rotated by the same amount (denoted here by $x$ and $y$) current passing in the $x$ direction generates a longitudinal voltage $V_x$ proportional to $\rho_a+\rho_c$, and a transverse voltage $V_y$ proportional to $\rho_c-\rho_a$.
			Simultaneous measurement of these voltages allows for both components of the resistivity to be measured simultaneously \cite{Walmsley2016}.
			(b) Temperature derivative of $\rho_c$ for each composition shown in (a).
			The dips marked by arrows correspond to the first (black) and second (red) CDW transitions in the parent compound, which are shown here to be suppressed towards zero as $x$ increases.
			We define the crossover temperature $T_{CDW}$ as the temperature at which the derivative reaches a local minimum.
			Curves were smoothed using a Loess filter prior to taking the numerical derivative.			
		}
		\label{fig:res}
	\end{figure}

	\Cref{fig:res} shows the in-plane components of the resistivity tensor of Pd$_x$ErTe$_3$, normalized to the room temperature value.
	For the parent compound, the resistivity is fourfold symmetric in the plane at room temperature, with anisotropy setting in sharply below the first CDW transition.\cite{Walmsley2016}
	Below the second CDW transition, the anisotropy decreases again.
	The two CDW transitions can be identified by the clear dips observed in the derivative of the $c$-axis resistivity.
	For $x\ne 0$, while the dips no longer signal true phase transitions, we can still use the evolution of the corresponding feature as $x$ increases to map out the crossover.
	Tracking these features across the composition series clearly shows that both CDW crossovers are suppressed by the inclusion of Pd in the crystals.
	The residual resistivity ratio also quickly drops with the introduction of Pd, as should be expected with increasing impurity scattering.

\begin{figure}
		\includegraphics[width=\columnwidth]{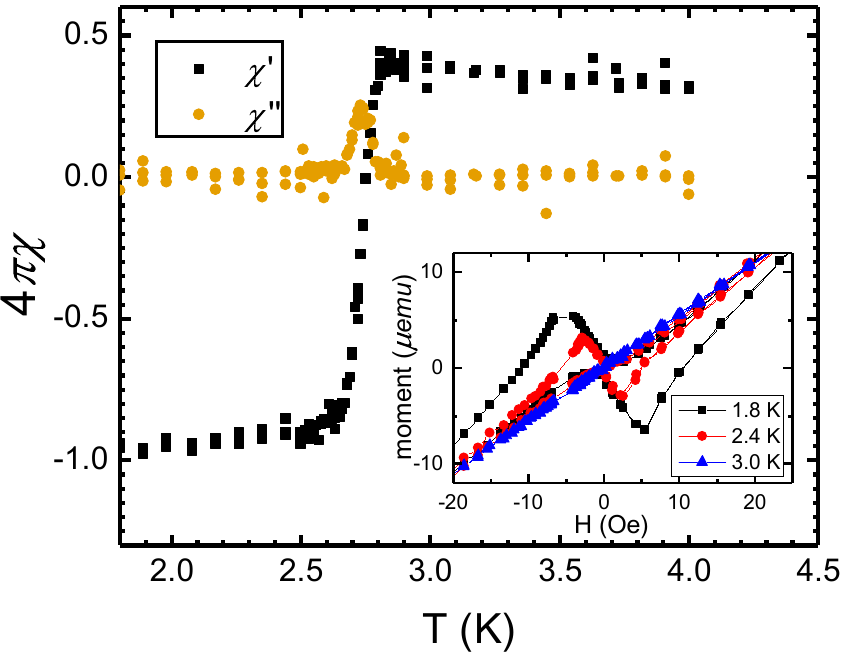}
		\caption{
			Representative AC magnetic susceptibility measurement for an $x=0.043$ sample.
			Field was applied parallel to the plane of the sample in this measurement.
			The sample shows near-perfect diamagnetism below $T_c=2.7~K$, suggesting that the superconducting state is bulk, rather than filamentary.
			$T_c$ is extracted as the peak in the imaginary part of the susceptibility.
			Inset: M(H) curves above and below $T_c$, showing the small $H_{c1}\approx 5~Oe$.
			The normal state just above $T_c$ also shows significant paramagnetism, a consequence of the proximity to antiferromagnetic ordering of the rare earth moments.
		}
		\label{fig:mag}
	\end{figure}

	The presence of a superconducting state at low temperatures is clearly detected by SQUID magnetometry, as shown in \cref{fig:mag}.
	Superconductivity first appears abruptly above our base temperature of 1.8~K when $x\approx0.02$, where the higher CDW crossover has been suppressed to $\approx$100~K, and remains near 2.5~K for all samples with $x>0.02$.
	Analysis of the superconducting volume fraction from zero-field cooled $M(T)$ curves shows that this is a bulk effect.
	The resistivity also drops to zero at the same temperature as the onset of the Meissner effect, within slight differences in thermometry in different cryostats.
	$H_{c1}$ determined from $M(H)$ hysteresis curves are on the order of 10-20~G.
	The area within these hysteresis loops is quite small, suggesting that vortex pinning is weak in this material, even at the largest $x$ values.
	However, it should be noticed that this signal is complicated by the presence of antiferromagnetic order of the localized rare-earth moments\cite{Ru2008, Deguchi2009} which produces a large background in the susceptibility measurement.
	
\begin{figure}
		\includegraphics[width=\columnwidth]{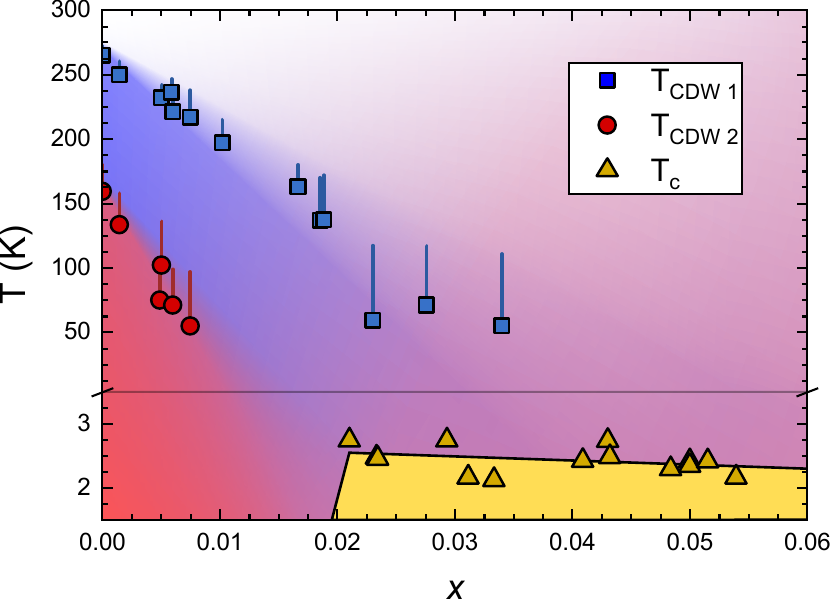}
		\caption{
			Phase diagram constructed from transport and magnetometry measurements.
			The temperature axis is rescaled below 3.5~K to better display the trend in superconducting transition temperatures.
			The second CDW is suppressed quickly as $x$ increases, and extrapolates to zero around $x=0.01$.
			A superconducting ground state emerges abruptly from below our base temperature of 1.8~K at $x = 0.02$, and continues with an approximately constant transition temperature out to the solubility limit of Pd.
			The first CDW extrapolates to zero around $x=0.035$.
			It should be noted that both CDW features decrease in magnitude and become harder to distinguish as $x$ increases.
			Lines extend upward from each data point to the first onset of the CDW signature in resistivity, illustrating how disorder smears the crossover.
			Background shading approximates the magnitude of CDW correlations along the two in-plane axes: blue signifies the primary $c$-axis CDW, and red signifies the $a$-axis.
			As disorder increases, the material begins to exhibit short range correlations along both directions (indicated by purple shading).
			True long-range-ordered CDW phases exist only on the $x=0$ axis.
		}
		\label{fig:pd}
	\end{figure}

	Resistivity and magnetometry measurements across the composition series are combined in \cref{fig:pd}.
	The resistivity signatures of both CDW crossovers disappear quickly, and the lower CDW is undetectable in resistivity above $x=0.01$.
	The upper CDW becomes undetectable above $x=0.035$.
	Unlike other perturbations such as chemical or hydrostatic pressure, under which one CDW is stabilized at the expense of the other\cite{Ru2008b, Hamlin2009}, we see that both CDWs are suppressed monotonically as $x$ increases.
	This is further evidence that the primary effect of the Pd intercalation is to introduce disorder rather than electronic doping.
	This demonstrates that Pd$_x$ErTe$_3$ is a promising model system for studying the interplay of incommensurate CDWs and disorder.
	
	While the phase diagram presented here appears at first glance to fit the paradigm of a superconducting dome around a quantum critical point, we emphasize that no true phase transition is being suppressed to zero temperature.
	Also, unlike other systems with putative quantum critical points, $T_c$ is never observed to decrease back to zero for larger $x$.
	A superconducting state with similar behavior is also observed for RTe$_3$ (R=Gd,Tb,Dy) under hydrostatic pressure\cite{Hamlin2009, Zocco2015}.
	However, there are a few notable differences between the responses of superconductivity in ErTe$_3$ to disorder and to pressure.
	First, the critical temperature achieved by disorder is universally lower than that achieved by the application of pressure.
	Ref. \onlinecite{Zocco2015} reported a trend of $T_c$ which closely matched that of elemental Te under pressure, and which reached approx. 4~K for pressures greater than 6~GPa, and tends to increase with pressure up to 14~GPa.
	In the presence of disorder, by contrast, $T_c$ is always below 3~K, and decreases slightly as $x$ increases.
	Both cases, however, exhibit a sharp increase in $T_c$ above a critical value of the tuning parameter.
	This is consistent with the interpretation that superconductivity and charge density waves compete over regions of the Fermi surface; as the CDW is suppressed, regardless of the tuning parameter, more electrons are available to condense into a superconducting state, which enhances $T_c$.
	
\begin{figure*}
		\includegraphics[width=\textwidth]{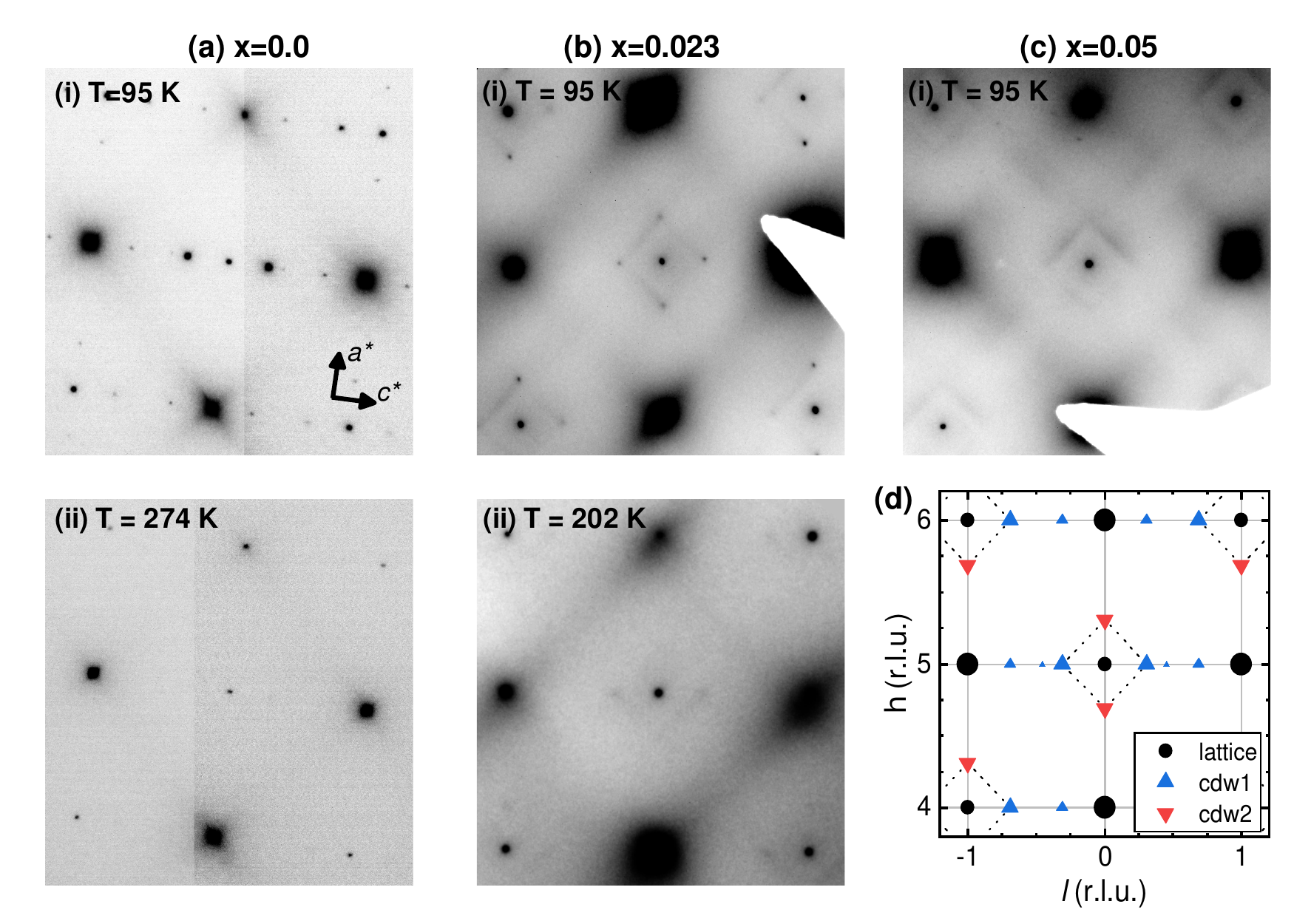}
		\caption{
			Representative selected area diffraction patterns (SADPs) at various Pd concentrations and temperatures.
			(a) Parent compound (ErTe$_3$) at 95~K (i) and 274~K $\sim T_{CDW,1}$ (ii), showing the disappearance of the satellite peaks on either side of the center peak at high temperatures.
			(b) SADP for an $x=0.023$ sample ($T_{CDW,1}\approx 97~K$) at 95~K (i), clearly showing satellite peaks forming along \emph{both} the $c^*$ and $a^*$ axes, despite the fact that the transport signature of the second CDW has been completely suppressed for this composition.
			This suggests that the introduction of disorder elicits a nearly 4-fold symmetric response from the material, reflective of the underlying electronic susceptibility and phonon dispersion.
			A small square of diffuse streaks also begins to appear between these peaks, suggestive of short-range ordering.
			It should be noted, however, that the width of the satellite peaks sets a lower bound on the correlation length of 40~nm, or about 94~unit cells.
			In (ii) remnants of the CDW peaks and the diffuse squares can be seen on this same sample even at 202~K, more than double the CDW crossover temperature.
			(c) Finally for a $x=0.05$, which shows no signatures of CDW in resistivity, no sharp satellite peaks are observed to the lowest temperature attainable, but the diffuse square remains visible at all temperatures.
			This is suggestive of a structural distortion with a similar $q$-vector, likely stemming from nucleation of CDW correlations by individual Pd atoms.
			Panel (d) shows a schematic describing the origin of the various peaks for the parent compound.
			The small upward triangles correspond to higher order CDW peaks which are only observable in (a).
		}
		\label{fig:tem}
	\end{figure*}

	In order to get a clearer picture of the structure of the CDW, we also conducted electron and x-ray diffraction measurements.
	In \cref{fig:tem} we show selected area electron diffraction patterns (SADPs) taken on $\approx$100~nm thick samples with varying concentrations.	
	As $x$ increases, two different trends develop in the SADPs.
	First, peaks appear at the expected CDW wavevector $q_{CDW,1}\approx0.29c^*$, but also along orthogonal directions far beyond the Pd concentrations where the second, CDW has been completely suppressed in resistivity.
	The $q$-vectors of peaks along the $a^*$ and $c^*$ axes are indistinguishable within the resolution of the images.
	Measurements of the diffraction pattern shows that the position of the CDW vector decreases by about 1.1\% of a reciprocal lattice unit (r.l.u.) between $x=0$ and $x=0.33$.

	As the intensity of the CDW peaks decreases with increasing $x$, the SADPs also begin to exhibit significant diffuse scattering indicative of short-range correlations, primarily along lines spanning between the two orthogonal CDW satellite peaks [dashed lines in \cref{fig:tem}(d)].
	While these patterns do become slightly brighter as the temperature decreases, the pattern was also observed at room temperature in both the $x=0.03$ and $x=0.04$ samples.
	This is indicative of a static structural distortion nucleated by the Pd impurities, which in heavily disordered samples persists even above the CDW transition temperature of the pristine samples, which is 270~K.
	
	The observed full-width at half maximum of the CDW satellite peaks in electron diffraction patterns is resolution limited, and corresponds to a minimum correlation length of approximately 40~nm.
	In the $x=0.03$ samples (the highest concentration for which peaks can be seen within the temperature range of the cryogenic holder), the CDW peaks are not detectably broader than in the clean limit.
	It should be noted that assuming the Pd atoms lie only between the Te bilayers, the average distance between Pd atoms is approximately 2~nm, much shorter than this lower bound on the correlation length.
	This seems to imply that the CDW is in the weak-pinning limit, where the potential of a single impurity is insufficient to locally lock the CDW phase.
	
	The apparent four-fold symmetry of the CDW peaks which appear in the electron diffraction patterns for the largest values of $x$ studied are consistent with the Friedel oscillations and associated lattice distortions that should be expected to arise from the calculated electronic susceptibility\cite{Johannes2008}.
	The diffuse streaks, which are present at all temperatures, qualitatively match the Lindhard susceptibility calculated in ref. \onlinecite{Johannes2008}, which exhibits peaks at both orthogonal CDW wavevectors, but also a ridge of local maxima along a line between them.
	In general, the distortion in the neighborhood of a delta-function impurity should be expected to produce the largest effect at the $q$-vectors for which the susceptibility is largest (and hence the lattice is softest), and these observations support this interpretation.
	A similar structure has also been observed in calculations of the energy-integrated phonon linewidth\cite{Maschek2018}, which suggests that phonons couple strongly to the electrons at the wavevectors of these diffuse streaks.
	This coupling does not, however, appear for the phonons responsible for the CDW order, which implies that the strong electron-phonon coupling at the streak wavevectors, and therefore the static distortions in the presence of impurities, involves phonons of different symmetries.
		
\begin{figure}
		\includegraphics[width=\columnwidth]{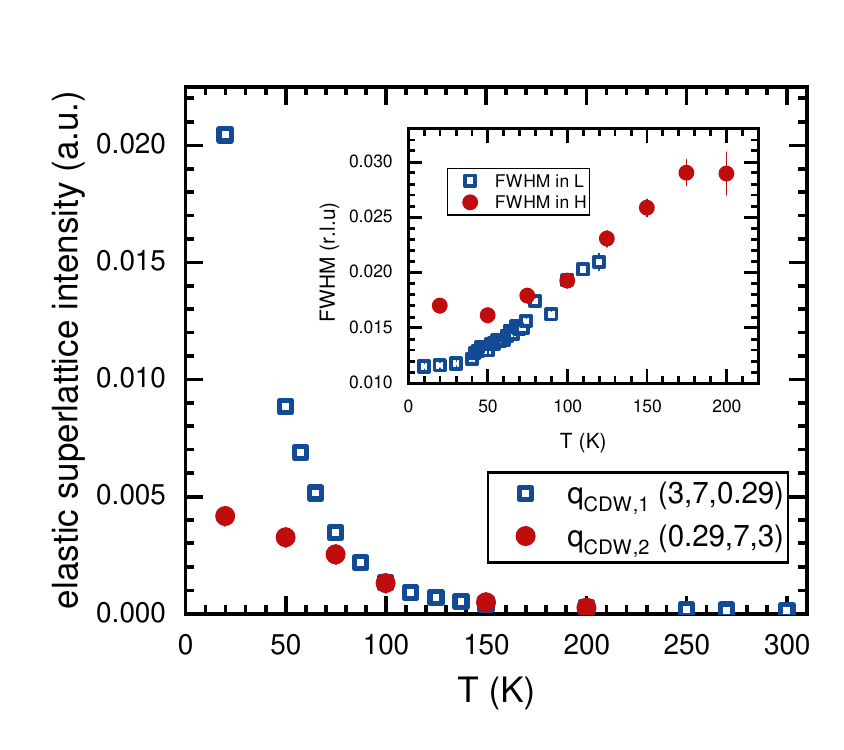}
		\caption{
			Measurement of the elastic CDW superlattice peak intensity observed in x-ray diffraction at zero energy transfer as a function of temperature for a sample with $x=0.023$.
			Data was taken at the HERIX beamline 30-ID at the Advanced Photon Source, with photon energy 23.72~keV and energy resolution of 1.5~meV.
			Above the transition temperature (95~K from resistivity) the amplitude of CDW correlations begin to grow identically, but the intensity of the primary CDW grows further at lower temperatures.
			Inset: FWHM measurements of longitudinal scans through $q_{CDW,1}$ and $q_{CDW,2}$ also show similar behavior above the crossover.
			Below the crossover, the correlation length of the secondary CDW along $a$ reaches a maximum around 90~K, while the correlation length of the primary continues to grow at least to the resolution limit.
			We estimate a reciprocal-space resolution for this beamline of 0.0112~r.l.u.
		}
		\label{fig:elastic}
	\end{figure}

	In order to characterize the satellite peak intensities down to lower temperatures, we also present energy-resolved x-ray scattering measurements on a sample with $x=0.023$.
	We focus on two superlattice peak positions, $(3, 7, 0.29)$, and $(0.29,7, 3)$, corresponding to the two CDW ordering vectors along the $c$- and $a$-axis, respectively.
	This composition only shows evidence of one CDW crossover in resistivity.
	\Cref{fig:elastic} shows the intensity of elastic scattering (zero energy transfer within resolution of 1.5 meV) at each of the two CDW wavevectors.
	Approaching the crossover temperature from above, both peaks grow in intensity at the same rate, while below the crossover (approx. 95~K for this composition) the two intensities differ.
	The peak corresponding to the primary CDW wavevector in the parent compound continues to become both stronger and sharper as temperature decreases, until the observed width levels off at the resolution limit of 0.0112 r.l.u. at 30~K.
	The intensity of the peak at the secondary CDW wavevector grows more slowly and the temperature dependence of the width levels of at approximately 90~K.
	Similar conclusions have recently been reached by STM measurements of Pd$_x$ErTe$_3$.\cite{Fang2019}

	Previous measurements on pristine ErTe$_3$\cite{Ru2008b} have demonstrated that the integrated intensity of the peak below $T_{CDW,1}$ follows a BCS temperature dependence with a sharp increase at the transition.
	The gradual increase of the intensity of both peaks for the intercalated sample is consistent with the CDW transition being smeared into a crossover in the presence of disorder.
	Consistent with the weak-pinning picture described above, the full width at half maximum (FWHM) of both peaks becomes sharper as the temperature decreases.
	While the $c$ axis correlations reach the resolution limit, for the CDW correlations along the $a$ axis we deduce a maximum possible correlation length of approx. 108 nm ($\approx25$ unit cells).
	This difference in correlation lengths confirms that the single crossover observed in resistivity for this composition still reflects the $c$-axis character of the first CDW transition in pristine ErTe$_3$, despite the slight enhancement of CDW correlations along the orthogonal direction.
	
	One possible explanation of the appearance of CDW correlations along both in-plane directions in TEM is that the Pd atoms nucleate stacking faults, disrupting the glide plane symmetry in the parent compound.
	To test this theory, the relative intensity of the symmetry allowed (0,6,1) peak and its forbidden counterpart (1,6,0) were compared in single-crystal x-ray diffraction.
	In the limit of a high density of stacking faults, one would expect that peaks of comparable magnitudes would be observed in both locations.
	In all crystals studied, only one peak was observed above the instrument noise floor, which gives confidence that the average distance between stacking faults is greater than the x-ray penetration depth (approx. \SI{2}{\micro\meter}).
	Further, crystals were then repeatedly exfoliated \textit{in situ} using adhesive tape, removing between \SI{4}{\micro \meter} and \SI{30}{\micro\meter} of material each time.
	The newly exposed surfaces always exhibited the same orientation in further scans.
	This sets a lower bound on the average distance between stacking faults which compares favorably to the thicknesses of samples used in TEM (of order $100$~nm) and transport (tens of microns).
	With this result, corroborated by the observed asymmetry in \cref{fig:elastic} of the two CDW orientations, we are confident that the effects reported here accurately represent the bulk behavior of the material.
	
	The appearance and evolution with temperature of superlattice peaks along both in-plane directions implies that the response of the material to disorder reflects the intrinsically pseudo-tetragonal nature of the electronic susceptibility and the phonon spectrum, as mentioned in the introduction.
	This raises the intriguing possibility that in-plane antisymmetric strain could be used to tune the system to an effectively tetragonal state, in which a vestigial nematic phase transition could be observed.
	(It should be noted that while the glide plane, being a nonsymmorphic symmetry operation, is fundamentally different from an in-plane strain, they both break the same point group symmetries as the vestigial nematic state.)
	Secondly, the mechanism of the relatively sharp increase of the superconducting transition temperature from below 1.8~K to 2.5~K remains in question.
	It may be that the Pd atoms increase the density of topological defects in the CDW order parameter, or that the magnitude of the CDW gap itself decreases as a consequence of disorder.
	Either mechanism would increase the density of states available at the Fermi level for superconducting pairing.
	Further studies are underway to clarify these details.

\section{Conclusions}
	We have presented the results of transport, magnetization, and scattering measurements on the model CDW system ErTe$_3$ intercalated with Pd.
	We have demonstrated that the primary effect of Pd intercalation is to provide a weak, random potential, which serves to suppress and smear the two CDW transitions into crossovers, and eventually elicit superconductivity.
	Our data also provides evidence for localized distortions of the crystal structure, which is strongest at the CDW wavevectors along both directions, as a consequence of the nearly 4-fold symmetric nature of the material.
	This work establishes Pd$_x$ErTe$_3$ as an easily tunable and weakly-coupled model system for the study of incommensurate, unidirectional CDWs in the presence of disorder.
	Study of this model system is a promising avenue for understanding the role of charge order within the complicated phase diagrams of strongly correlated systems such as the cuprate superconductors.
		
\section{Acknowledgments}
	The authors wish to thank Phil Walmsley, Dale Burns, and Ann Marshall for experimental assistance.
	We also thank Steven Kivelson, Laimei Nie, and Alexander Hristov for fruitful discussions.
	This work was supported by AFOSR Grant No. FA9550-09-1-0583.
	Part of this work was performed at the Stanford Nano Shared Facilities (SNSF) supported by the National Science Foundation under award ECCS-1542152.
	This research used resources of the Advanced Photon Source, a U.S. Department of Energy (DOE) Office of Science User Facility operated for the DOE Office of Science by Argonne National Laboratory under Contract No. DE-AC02-06CH11357.
	J.S. acknowledges support as an ABB Stanford Graduate Fellow.
	F.W. was supported by the Helmholtz Association under contract VH-NG-840.
	S.R. was supported by the U.S. DOE, Office of Science, Materials Sciences and Engineering Division.

	\bibliographystyle{apsrev4-1}
\end{document}